# THE ORIGIN OF DARK MATTER AND THE COSMOLOGICAL CONSTANT

Gerald E. Marsh
Argonne National Laboratory (Ret)
gemarsh@uchicago.edu

**ABSTRACT**

**The concept of exotic charged dust is introduced here to represent dark matter. The term "exotic" means that the dust is not composed of normal matter, and the charge–for lack of a better term–is not an electric charge. It is also shown that the often-used approximate expression for dark matter density corresponds to an exact solution of the coupled Einstein-Maxwell equations for charged dust. The Einstein field equations tells us is that for a flat universe the cosmological constant acts as a gravitating negative energy distribution that is gravitationally repulsive; following Schwinger, it is shown that the vacuum must have a negative energy spectrum if QFT is to be gauge invariant.**

The time from the beginning of the universe, what Fred Hoyle derogatively called the "big bang ", to the Planck time at $10^{-44}s$ is often called the Planck epoch. During the period from the beginning until about $10^{-44}s$ the four fundamental forces of nature—gravity, electromagnetism, and weak and strong nuclear interactions—are thought to have been unified into a single force. During this period the universe was thought to have had perfect symmetry, although the group associated with such symmetry is unknown. At ~$10^{-44}s$ spontaneous symmetry breaking led to the separation of the gravitational force from the other three forces. Figure 1 shows the time sequence of symmetry breaking and the associated groups.

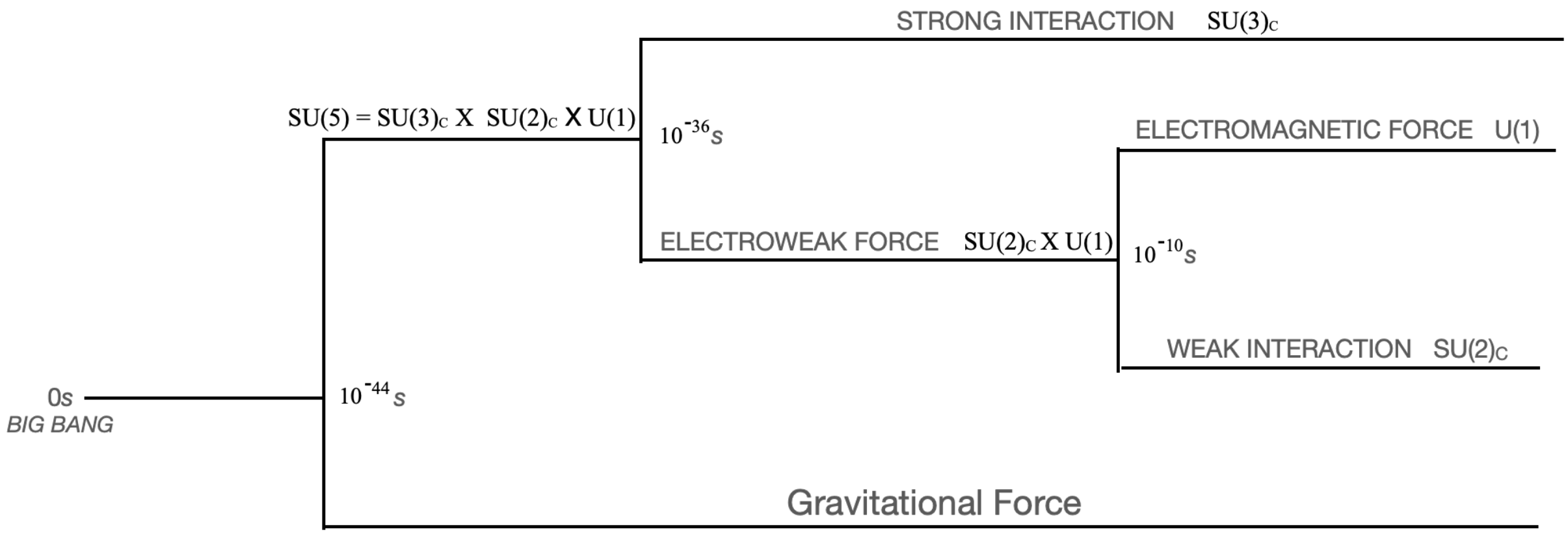

Figure 1. The approximate time sequence from the origin of the universe showing the spontaneous symmetry breaking times along with their associated groups. The vertical lines correspond to the times of symmetry breaking. After $10^{-36}s$, quantum field theory as well as quantum chromodynamics becomes relevant.

In Fig. 1, the conventional choice of SU(5) of the Georgi-Glashov model is used to represent the group for grand unification (without gravity) after $10^{-44}s$. The are several reasons, however, why SU(5) seems inappropriate including the fact that it predicts many unobserved phenomena such as proton decay and other baryon and lepton number violating processes. The actual group remains unknown.

It is often said that it is necessary to have a theory of quantum gravity in order to successfully unify gravity with the other forces of nature. And while both quantum field theory and general relativity have had spectacular success in explaining the domains of their applicability, there is no experimental evidence that there is a close relationship between the two. There is no evidence whatsoever that gravitational waves are quantized. All claims that a relationship exists are based on theoretical expectations. In the end, it may not be possible to quantize gravity as given by the general theory of relativity.

The nature of the vacuum comes into play after the symmetry breaking at $10^{-44}s$ when gravity separates from the other forces of nature. At that time, the crucial question is whether or not negative energy states exist for the vacuum. Since the universe remains flat from at least $10^{-36}s$, there must be a cosmological constant that acts as a negative energy density. This is discussed in the section below titled: "Perfect Symmetry Breaking at $10^{-44}$s, Dark Matter, and the Cosmological Constant".

Many have thought that the vacuum has an intrinsic energy density. The evidence given to support the reality of the various possible contributions to the vacuum energy is the Casimir effect, which is a consequence of the lowest order vacuum fluctuations, and higher order effects like the Lamb shift. But there are alternative explanations that could explain the Casimir effect. It could result from fluctuations associated with the constituents of the plates rather than vacuum fluctuations. Schwinger’s source theory takes this point of view and avoids vacuum fluctuations in both the Casimir and higher order QED effects. As Schwinger put it, ". . . the vacuum is not only the state of minimum energy, it is the state of zero energy, zero momentum, zero angular momentum, zero charge, zero whatever." Pauli also seemed to agree with this position when commenting on field fluctuations in quantum field theory, ". . . it is quite impossible to decide whether the field fluctuations are already present in empty space or only created by the test bodies." Sometime after the Schwinger quote above, Schwinger showed that negative energy states must exist if QFT is to be gauge invariant, which in turn means that a term (called the Schwinger term) must vanish. He then showed that if the term does vanish, so that QFT is gauge invariant, the vacuum state couldn’t be the state with lowest field energy. As put by Schwinger,[1] ". . . it is customary to assert that the electric charge density of a Dirac field commutes with the current density at equal times, since the current vector is a gauge-invariant bilinear combination of the Dirac fields. It follows from the conservation of charge that the charge density and its time derivative, referring to any pair of spatial points at a common time, are commutative. But this is impossible if a lowest energy state—the vacuum—is to exist."

Schwinger's finding is crucial for the existence of a cosmological constant in Einstein's gravitational field equations. Since his argument is rather opaque in his 1959 *Physical Review Letters* article, here is a derivation:

The Schwinger term (ST) is given by

$$ST(\vec{y},\vec{x}) = \langle\left[\hat{\rho}(\vec{y}),\hat{\vec{J}}(\vec{x})\right]\rangle \tag{1}$$

Taking the divergence of the Schwinger term and using the relation

$$\mathrm{i}\left[\hat{\mathrm{H}}_0,\,\hat{\rho}(\vec{x})\right] \;=\; -\,\nabla\cdot\,\hat{\vec{J}}(\vec{x}), \tag{2}$$

where $\hat{\mathrm{H}}_0$ is the free-field Hamiltonian when the electromagnetic 4-potential vanishes, results in

$$\nabla_{\vec{x}}\cdot\left[\hat{\rho}(\vec{y}),\,\hat{\vec{J}}(\vec{x})\right] \;=\; \left[\hat{\rho}(\vec{y}),\,\nabla\cdot\hat{\vec{J}}(\vec{x})\right] \;=\; -\mathrm{i}\left[\hat{\rho}(\vec{y}),\,\left[\hat{\mathrm{H}}_0,\,\hat{\rho}(\vec{x})\right]\right]. \tag{3}$$

Expanding the commutator on the right-hand side of Eq. (3) yields the vacuum expectation value

$$i\nabla_{\vec{x}}\cdot\left\langle 0\left|\left[\hat{\rho}(\vec{y}),\hat{\vec{J}}(\vec{x})\right]\right|0\right\rangle = -\langle 0|\,\hat{\mathrm{H}}_0\,\hat{\rho}(\vec{x})\hat{\rho}(\vec{y})\,|0\rangle +$$
$$\langle 0|\;\hat{\rho}(\vec{x})\hat{\mathrm{H}}_0\hat{\rho}(\vec{y})\,|0\rangle + \langle 0|\;\hat{\rho}(\vec{y})\hat{\mathrm{H}}_0\hat{\rho}(\vec{x})\,|0\rangle - \langle 0|\;\hat{\rho}(\vec{y})\hat{\rho}(\vec{x})\hat{\mathrm{H}}_0\,|0\rangle.$$

(4)

It is here that one makes the assumption that the vacuum is the lowest energy state. This done by writing $\hat{\mathrm{H}}_0$ |0> = <0| $\hat{\mathrm{H}}_0$ = 0. As a result, Eq. (4) may be written as

$$i\nabla_{\vec{x}}\cdot\left\langle 0\left|\left[\hat{\rho}(\vec{y}),\hat{\vec{J}}(\vec{x})\right]\right|0\right\rangle = \langle 0|\;\hat{\rho}(\vec{x})\hat{\mathrm{H}}_0\hat{\rho}(\vec{y})\,|0\rangle +$$
$$\langle 0|\;\hat{\rho}(\vec{y})\hat{\mathrm{H}}_0\hat{\rho}(\vec{x})\,|0\rangle.$$

(5)

Multiply both sides of the last equation by *f*(*x*)*f*(*y*) and integrate over *x* and *y*. The right-hand side of Eq. (5) becomes

$$\int d\vec{x}\,d\vec{y}\,\left\{\left\langle 0\right|f(\vec{x})\hat{\rho}(\vec{x})\hat{\mathrm{H}}_0 f(\vec{y})\hat{\rho}(\vec{y})\left|0\right\rangle + \left\langle 0\right|f(\vec{y})\hat{\rho}(\vec{y})\hat{\mathrm{H}}_0 f(\vec{x})\hat{\rho}(\vec{x})\left|0\right\rangle\right\}.$$

(6)

If Schwinger's "arbitrary linear functional of the charge density" is defined as

$$F = \int f(\vec{x})\hat{\rho}(\vec{x})d\vec{x} = \int f(\vec{y})\hat{\rho}(\vec{y})d\vec{y} \;\;,$$

(7)

the right-hand side of Eq. (5) becomes

$$2\left\langle 0\right|F\hat{\mathrm{H}}_0 F\left|0\right\rangle = 2\sum_{m,n}\langle 0|\,F\,|m\rangle\left\langle m\right|\hat{\mathrm{H}}_0\left|n\right\rangle\langle n|\,F\,|0\rangle =$$
$$2\sum_n E_n\langle 0|\,F\,|n\rangle\langle n|\,F\,|0\rangle = 2\sum_n E_n|\,\langle 0|\,F\,|n\rangle\,|^2 > 0.$$

(8)

The left-hand side of Eq. (8)—essentially the form used by Schwinger—is here expanded to explicitly show the non-vanishing matrix elements between the vacuum and the other states of necessarily positive energy. This shows that if the vacuum is assumed to be the lowest energy state, the Schwinger term cannot vanish, and the theory is not gauge invariant.

For the sake of completeness, it is readily shown that the left side of Eq. (5) becomes

$$i\int \nabla_{\vec{x}}\cdot\left\langle 0 \left| \left[\hat{\rho}(\vec{y}), \hat{\vec{J}}(\vec{x})\right] \right| 0 \right\rangle f(\vec{x})f(\vec{y})\mathrm{d}\vec{x}\mathrm{d}\vec{y} \;=\; i\,\langle 0\,|\,[\partial_t F, F]\,|0\rangle, \tag{9}$$

so that combining Eqs. (8) and (9) yields a somewhat more explicit form of the result given by Schwinger,

$$i\,\langle 0\,|\,[\partial_t F, F]\,|0\rangle \;=\; 2\sum_n E_n |\langle 0| F |n\rangle|^2 > 0. \tag{10}$$

Because the Schwinger term does not vanish, if QFT is to be gauge invariant the vacuum must have a negative energy spectrum.

## Dark Matter[2]

The standard ΛCDM model has as its principal matter component collisionless cold dark matter of an unknown nature. The rotation curves of spiral galaxies as well as the inferred mass of galaxy clusters are best explained by the existence of dark matter that dominates their mass content. Relatively recent work on colliding galaxy clusters appear to confirm this supposition.[3,4] Other possibilities, such as modifying Newton's equations of motion or postulating a change in the gravitational interaction between dark and normal matter have been proposed, but have not gained favor. The history of the dark matter problem can be found in the *Nature Review Article* by de Swart, Bertone, and van Dongen.[5]

In the case of the rotation curves of galaxies, the density distribution of dark matter is generally assumed to be spherical and to have an isothermal equation of state; i.e., a polytropic equation of state ($P = K\rho^{\gamma}$) where $\gamma = 1$. The hydrostatic balance equation may then be integrated to yield

$$\rho = \rho_0 \, exp(-\Phi/K). \tag{11}$$

where $\Phi$ is the gravitational potential. $\Phi/K$ must then be a solution of the isothermal Lane-Emden equation. Non-singular solutions can be obtained by imposing appropriate boundary conditions, such as requiring that the solution and its first derivative vanish at the origin. The result is an exponential solution for the density of the form

$$\rho(r) = \frac{\rho_0}{exp(\Phi/K)}. \tag{12}$$

The isothermal Lane-Emden equation cannot be solved analytically and consequently $\Phi/K$ is expanded in a power series. The requirement that the first derivative vanish at the origin limits the expansion to even powers starting with $(\Phi/K)^2$. Expanding the exponential in the denominator of Eq. (12), keeping only the first two terms, and using the coefficient given by Chandrasekhar[6] for the leading $(\Phi/K)^2$ term results in the often-used expression for the dark matter density,

$$\rho(r) = \rho_0 \frac{{r_0}^2}{{r_0}^2 + r^2}. \tag{13}$$

where $r_0 = \sqrt{6K/4\pi G\rho_0}$. It will be shown in the next section that the right-hand side of this approximate expression corresponds to an exact solution of the coupled Einstein-Maxwell equations for charged dust.

**Charged Dust**

The form of the metric for charged dust was introduced by Majumdar[7] and Papapetrou[8]. It is spherically symmetric and static, and can be motivated by considering the Reissner-Nordström metric

$$ds^2 = \left(1 - \frac{2m}{r} + \frac{Q^2}{r^2}\right)dt^2 - \left(1 - \frac{2m}{r} + \frac{Q^2}{r^2}\right)^{-1} dr^2 - dr^2(d\theta^2 + \sin^2\theta\, d\phi^2). \tag{14}$$

Assume the extreme form of this metric where $|Q| = m$, and introduce the isotropic coordinates $\bar{r} = r - m$. Doing so results in the metric (dropping the bar above $r$)

$$ds^2 = f^2 dt^2 - f^{-2}[dr^2 + r^2(d\theta^2 + \sin^2\theta\, d\phi^2)]. \tag{15}$$

where $f = \left(1 + \frac{m}{r}\right)^{-1}$ and it is understood that isotropic coordinates are used in what follows.

Using Newtonian mechanics and classical electrostatics, it is straightforward to show that a system of charged particles of mass $m_i$and charge $q_i$, where all of the particles have the same sign charge, will be in static equilibrium if $||q_i| = G^{1/2} m_i$. For a continuous distribution of mass $\rho$ and charge $\sigma$, there will be equilibrium everywhere if $|\sigma| = G^{1/2}\rho$. This is what is known as charged dust. It has a general relativistic analog that was discovered by Papapetrou and Majumdar. Although spatial symmetry is not required, spherical symmetry will be assumed here.

Note that the extremal condition $q = G^{1/2} m$ means that if $q$ is chosen to be the minimal charge of one electron or $2.9 \times 10^{-19}$ coulomb, then there is a minimal mass of $3.5 \times 10^{-13}$ kilogram. This minimal mass is about $10^4$ less than the value to the reduced Plank mass of $\sqrt{\frac{\hbar c}{8\pi G}} = 4.3 \times 10^{-9}$ kilogram. Converting the minimal mass into energy in GeV gives $2 \times 10^{14}$ GeV.

At $10^{-36}$s after the time of the creation of the universe (see Fig. 1) the temperature was ~$10^{28}$ K, which–using $E = kT$–corresponds to an energy of ~$10^{15}$ Gev, close to the minimal mass energy.

The equilibrium of charged dust in general relativity has been treated extensively by W.B. Bonnor and others since the early 1960s. It is his paper on the equilibrium of a charged sphere[9] that forms the embarkation point for the work here.[10] The Einstein and Maxwell field equations applied to the metric of Eq. (15) show that the Newtonian condition for equilibrium given above must also hold in general relativity. In what follows, the charge will be chosen to be positive.

Bonnor obtained the equation that relates the general form of $f$ to the density,

$$ff'' - 2f'^2 + \frac{2}{r}ff' - 4\pi\rho = 0. \tag{16}$$

Unfortunately, this equation is completely intractable unless $\rho = 0$, and as a result one is reduced to guessing a form for the function $f$ and hoping that the equation yields a physically meaningful density distribution. The question then reduces to whether it is possible to find a function $f(r)$ that would result in a radially unlimited density distribution matching that given in Eq. (13) for dark matter. It turns out that the substitution of

$$f(r) = \sqrt{\frac{4}{3}\pi\rho_0}\ (a^2 + r^2)^{1/2} \tag{17}$$

into Eq (16) yields

$$\rho(r) = \rho_0 \frac{a^2}{a^2 + r^2} \tag{18}$$

where $a$ is now a free constant. This has the same form as Eq. (14) except that now the equality is exact and $\rho(r)$ is derived from a solution of the Einstein-Maxwell field equations. This is somewhat surprising given that the origins of Eq. (13) and Eqs. (17) and (18) are so different.

If one tries to generalize the solution given by Eq. (17) to

$$f(r) = \sqrt{\frac{4}{3}\pi\rho_0}\ (a^\gamma + r^\gamma)^{1/2} \tag{19}$$

the solution for the density is found to be

$$\rho(r) = \rho_0 \frac{r^{\gamma-2}\gamma[2a^\gamma(1+\gamma) - r^\gamma(\gamma-2)]}{12(a^\gamma + r^\gamma)}. \tag{20}$$

For $\gamma > 2$, the density has negative values for some range of $r$. Consequently, one must impose the condition that $\gamma \leq 2$. The plot of the density for $\gamma = 2$ and $\gamma < 2$ is shown in Fig. 2. The cusp in the density for $\gamma < 2$ is a result of the density being singular at $r = 0$. Singular density functions are sometimes used to model the mass distributions in elliptical galaxies. This will not be discussed here. Note that for $\gamma = 2$ Eq. (20) reduces to Eq. (18).

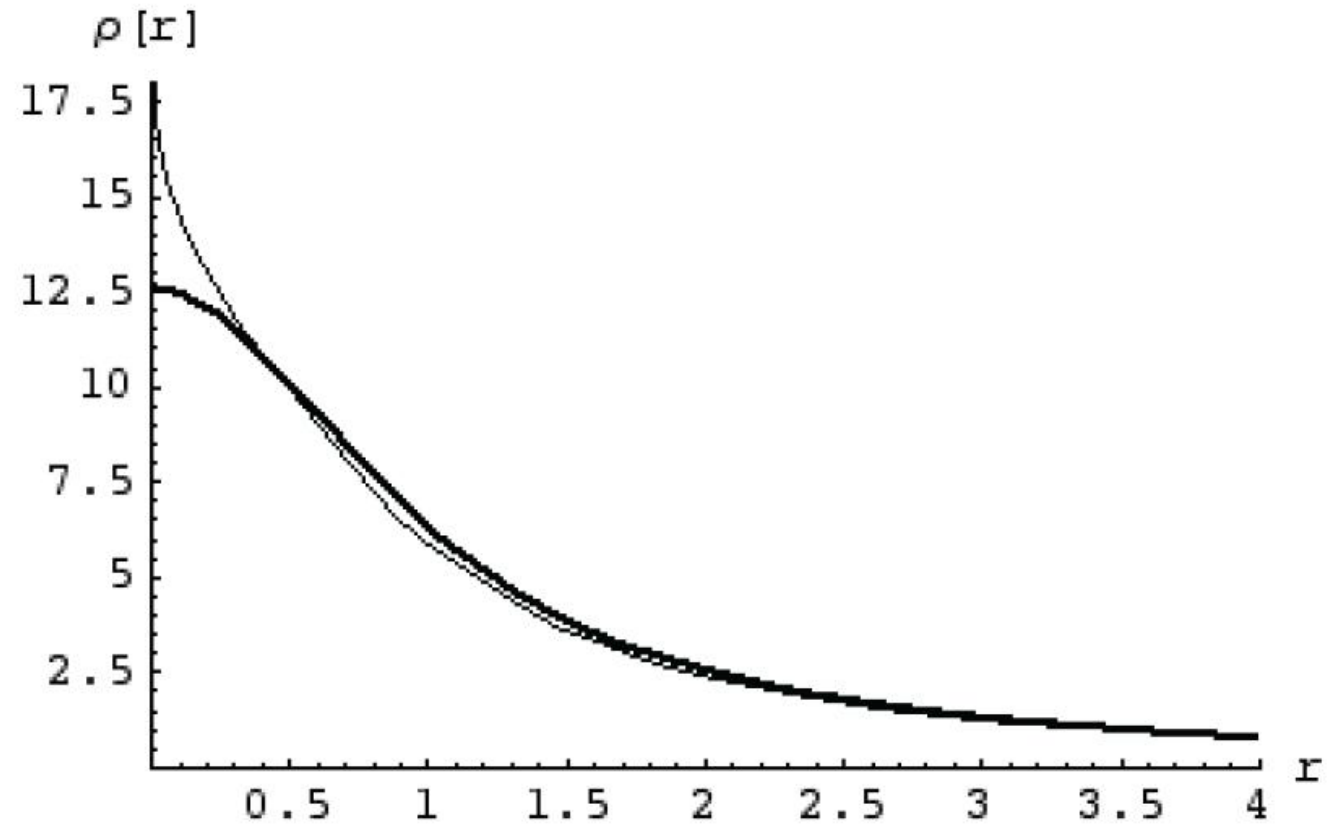

Figure 2. A plot of the density distribution as a function of r given by Eq. (20) for $\gamma = 2$ (dark) and $\gamma = 1.9$ (light) for $\rho_0 = a = 1$. The density for $\gamma = 1.9$ is singular at $r = 0$. It is therefore $\gamma = 2$ that is of interest here.

To summarize this section, it has been shown that the approximate solution to the isothermal Lane-Emden equation, often used–with suitable boundary conditions at the origin–to model dark matter halos, corresponds to an exact solution of the Einstein-Maxwell equations for charged dust.

**Perfect Symmetry Breaking at $10^{-44}$s, Dark Matter, and the Cosmological Constant**

Given what is currently known, it is arguable that the time of separation of gravity from the other forces of nature is the time when dark matter and the cosmological constant came into existence. And as one can see from Fig. 3, if one assumes an inflationary universe, it is also the time of the slow beginning of the rapid expansion (inflation) of the size of the universe.

Big bang cosmology predicts that large numbers of heavy stable monopoles would be produced in the very early universe. In the inflationary universe monopoles are allowed to exist if they were produced prior to the period of inflation when the cosmological constant and dark matter are assumed to have come into existence. The basis of this argument is the lack of observational evidence for monopoles; inflation would dilute the density of monopoles to agree with the lack of evidence for their existence. The argument would not be needed if the monopoles were composed of dark matter.

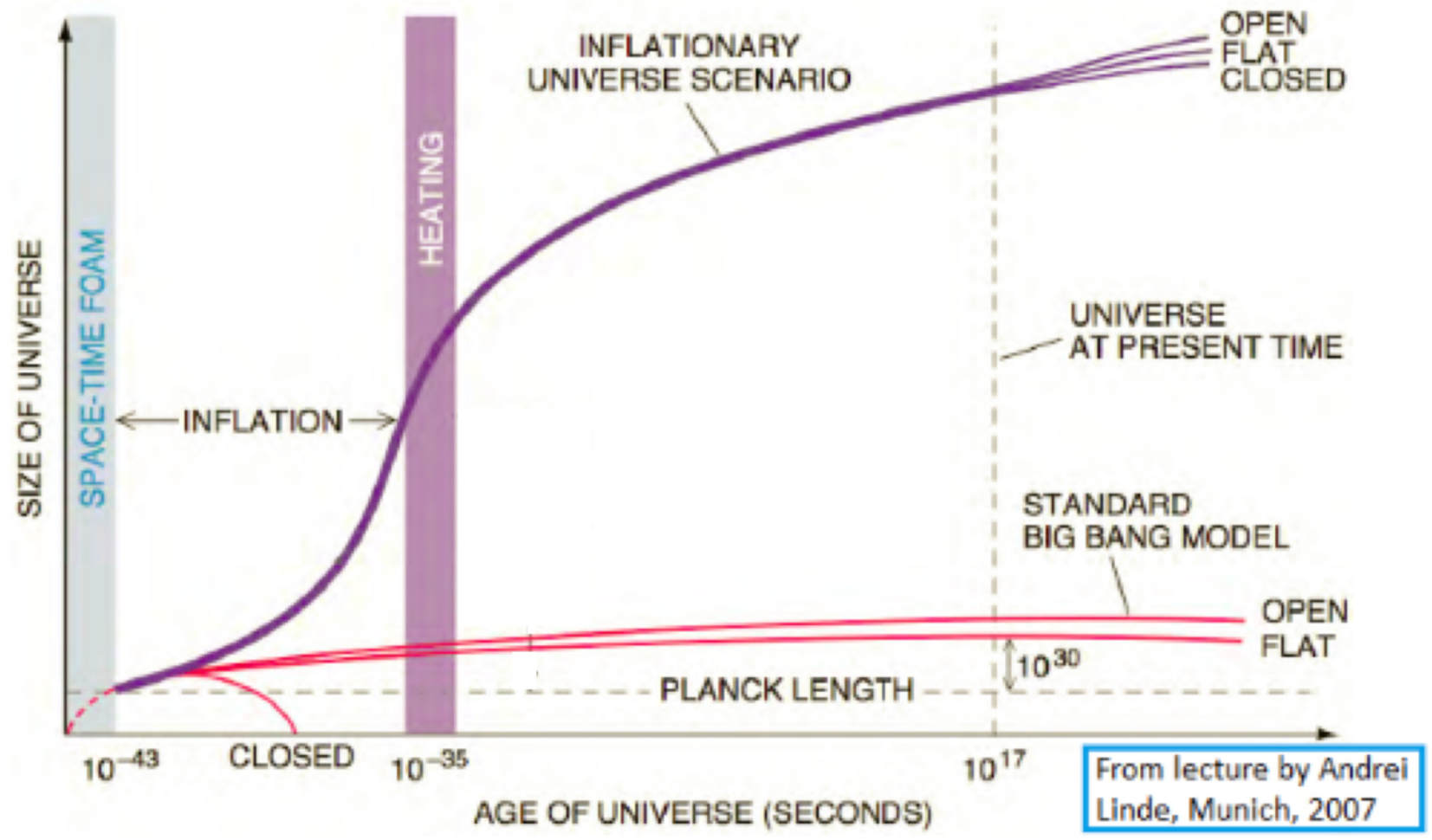

Figure 3. The size of the universe as a function of its age. The beginning of inflation corresponds to time of separation of gravity from the other forces of nature at the time of perfect symmetry breaking at $10^{-44}s$. [Adapted from the lecture figure.]

Observation tells us that the universe is essentially flat, although there is some controversial evidence of a continuing acceleration. What the Einstein field equations with the cosmological constant $\Lambda$ tells us is that for a flat universe the cosmological constant acts as a gravitating negative mass distribution. For the universe to be essentially flat, the cosmological constant must compensate for real mass and energy as well as positive mass dark matter since it acts gravitationally. As put by Einstein in 1918, "a modification of the theory is required such that 'empty space' takes on the role of gravitating negative masses which are distributed all over the interstellar space." Today one replaces the cosmological constant with a "dark energy" that causes a negative pressure. Referring to the WMAP (Wilkinson Microwave Anisotropy Probe) data, NASA puts it this way, the data "raise the possibility that the universe contains a bizarre form of matter or energy that is, in effect, gravitationally repulsive. The cosmological constant is an example of this type of energy". Moreover, the cosmological constant could not be considered as composed of repulsive particles since, in order to retain a constant density as the universe expands, there would have to be a continuous creation of such particles. Thus, the origin and nature of the cosmological constant remain mysterious. Considering the cosmological constant as due to the existence of negative energy states eliminates this problem.

**Exotic Charged Dust**

Bonnor and others, since the early 1960s, assumed that charged dust was composed of normal matter and electric charge. Use of the concept in cosmology for dark matter means there would have to be a way to retaining charges of one sign, without them being neutralized, throughout the universe. Here, the concept of exotic charged dust is introduced to solve this conundrum. The term "exotic" means that the dust is not composed of normal matter, and the charge–for lack of a better term–is not an electric charge. It is assumed that the exotic mass will continue to obey the usual equations of general relativity, and the exotic charge, Coulomb's law; the notation for mass and charge will also remain the same, but henceforth they mean exotic mass and charge.

And, as shown above, if one accepts exotic charged dust for the nature of dark matter, its distribution in galactic halos would agree with observation and also correspond to an exact solution of the Einstein-Maxwell equations.

**Cosmological models**

In cosmology, one often chooses the model of a perfect fluid so that the equation of state is closely related to that of an ideal gas. The equation of state is then characterized by a dimensionless number $w$, which is the ratio of the pressure $p$ to its energy density $\rho$; i.e., $w = p/\rho$. If the expansion of the universe is due to a type of scalar field $\Phi$, it can be characterized as a type of perfect fluid with the equation of state

$$w = \frac{\frac{1}{2}(\dot{\Phi})^2 - V(\Phi)}{\frac{1}{2}(\dot{\Phi})^2 + V(\Phi)},$$

(23)

where $V(\Phi)$ is the potential energy and $\Phi$ is a function of the spatial coordinates $x_i$. The importance of this equation is that for a free field, where $V = 0$, the equation of state parameter $w$ is +1, which is inconsistent with the cosmological observations and the cosmological constant Λ. On the other hand, if the kinetic energy of the field vanishes $w = -1$, which is equivalent to a cosmological constant Λ and also consistent with observations. Dark matter would be identified with static positive exotic charge and mass so that $w = -1$.

It has been assumed here that at the time of separation of gravity from the other forces of nature is the beginning of the time period when dark matter and the cosmological constant came into existence. In the section above on charged dust it was found that the extremal condition $q = G^{1/2}m$ means that if $q$ is chosen to be the minimal charge of one electron or 2.9 X $10^{-19}$ coulomb, then there is a minimal mass of 3.5 X $10^{-13}$ kilogram. Converting the minimal mass into energy in GeV gives 2 X $10^{14}$ GeV. At $10^{-36}$s after the time of the creation of the universe (see Fig. 1) the temperature was ~$10^{28}$ K, which–using $E = kT$–corresponds to an energy of ~$10^{15}$ Gev. This is consistent with the early beginning for the cosmological constant and dark matter.

At the time of the big bank, it is often supposed that the universe was initially a closed manifold whose curvature was later vastly reduced by the inflationary scenario. And while this is intuitively attractive, it may not be true. If initially the positive energy was balanced by a negative energy, later associated with the cosmological constant, the universe could well have been flat from the beginning and this would then also include the inflationary period.